\documentclass[%
 aip,
 amsmath,amssymb,
 reprint,%
]{revtex4-2}

\usepackage{graphicx}
\usepackage{dcolumn}
\usepackage{bm}

\usepackage[utf8]{inputenc}
\usepackage[T1]{fontenc}
\usepackage{mathptmx}
\usepackage{etoolbox}

\makeatletter
\def\@email#1#2{%
 \endgroup
 \patchcmd{\titleblock@produce}
  {\frontmatter@RRAPformat}
  {\frontmatter@RRAPformat{\produce@RRAP{*#1\href{mailto:#2}{#2}}}\frontmatter@RRAPformat}
  {}{}
}%
\makeatother
\begin{document}

\preprint{AIP/123-QED}

\title[Infrared super-resolution wide-field microscopy using sum-frequency generation]{Infrared super-resolution wide-field microscopy using sum-frequency generation}
\author{Richarda Niemann}
\author{Sören Wasserroth}%
\affiliation{ 
Fritz-Haber-Institut der Max-Planck-Gesellschaft, Faradayweg 4-6, 14195 Berlin, Germany
}%

\author{Guanyu Lu}
\affiliation{%
Vanderbilt University, Nashville, Tennessee 37235, USA
}%
\author{Sandy Gewinner}
\author{Marco De Pas}
\author{Wieland Schöllkopf}
\affiliation{ 
Fritz-Haber-Institut der Max-Planck-Gesellschaft, Faradayweg 4-6, 14195 Berlin, Germany
}%
\author{Joshua D. Caldwell}
\affiliation{%
Vanderbilt University, Nashville, Tennessee 37235, USA
}%
\author{Martin Wolf}
\affiliation{ 
Fritz-Haber-Institut der Max-Planck-Gesellschaft, Faradayweg 4-6, 14195 Berlin, Germany
}%
\author{Alexander Paarmann}
\affiliation{ 
Fritz-Haber-Institut der Max-Planck-Gesellschaft, Faradayweg 4-6, 14195 Berlin, Germany
}%

\date{\today}

\begin{abstract}
Super-resolution microscopy in the visible is an established powerful tool in several disciplines. In the infrared (IR) spectral range, however, no comparable schemes have been demonstrated to date. In this work, we experimentally demonstrate super-resolution microscopy in the IR range ($\lambda_{IR}\approx 10-12\,\mu$m) using IR-visible sum-frequency generation. We operate our microscope in a wide-field scheme and image localized surface phonon polaritons in 4H-SiC nanostructures as a proof-of-concept. With this technique, we demonstrate an enhanced spatial resolution of $\sim\lambda_{IR}/9$, enabling to resolve the polariton resonances in individual sub-diffractional nanostructures with sub-wavelength spacing. Furthermore we show, that this resolution allows to differentiate between spatial patterns associated with different polariton modes within individual nanostructures.

\end{abstract}

\maketitle


In classical light microscopy the imaging resolution is defined by the diffraction limit of light pointed out by Abbe \cite{Abbe1873}, leading to a spatial resolution limited by the imaging wavelength and the numerical aperture of the objective. Many techniques have been developed and demonstrated to overcome this diffraction limit, such as stimulated emission depletion (STED) microscopy\cite{Vicidomini2018}, structured illumination microscopy (SIM) \cite{Heintzmann2017}, and stochastic optical reconstruction microscopy (STORM).\cite{Bates2013} Additionally, nanotip-based approaches such as scattering-type scanning nearfield optical microscopy\cite{Heinzelmann1994} (s-SNOM) and photothermal-induced resonance (PTIR) microscopy\cite{Dazzi2009} were introduced. One attractive alternative to these established approaches is nonlinear-optical frequency mixing, such as sum-frequency generation (SFG).\cite{Shen2016} Here, the light at the resonant imaging wavelength is upconverted with a second, nonresonant light source at much shorter wavelength providing resonant contrast given by the former and high spatial resolution defined by the latter. 
    
\begin{figure}[h!]
\centering
\includegraphics[scale=1]{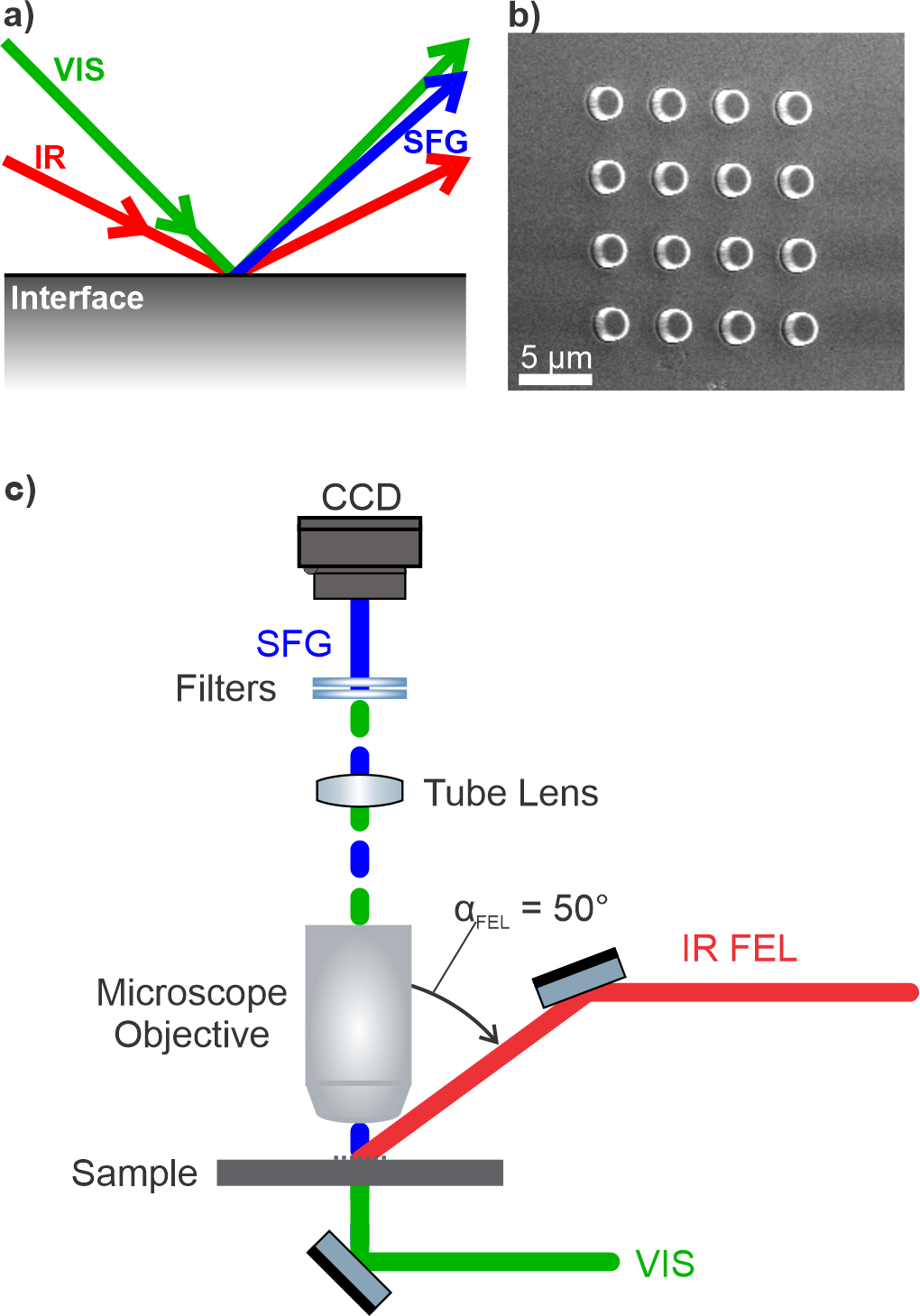} 
\caption{(a) SFG process. (b) SEM image of a 4$\times$4 nanopillar patch, consisting of NPs with a diameter of 2\,$\mu $m, a period of 5\,$\mu$m and a height of 1\,$\mu$m taken under an angle of 30\,$^\circ$. (c) Scheme of the microscope setup. The visible beam illuminates the sample from the back in normal incidence. The IR-FEL beam illuminates the sample from the front under an incident angle of 50$^\circ$. The SFG light is collected by the microscope consisting of a 50$\times$ objective and a 200\,mm tube lens. Several filters block the VIS light from entering the CCD camera chip. Here, only the SFG signal is finally imaged. Visualization of optical components taken from Ref. [\citenum{FranzenInk}]. }
\label{fig1}
\end{figure}

Specifically in infrared-visible (IR-VIS) SFG, two laser beams of frequencies $\omega_{IR}$ and $\omega_{VIS}$ are overlapped in a nonlinear material, which results in the emission of a third beam of a frequency $\omega_{SFG}=\omega_{IR}+\omega_{VIS}$. A schematic of the process is depicted in Fig.\,\ref{fig1}a). SFG is an even-order non-linear technique, which makes it a useful tool for studying materials with broken inversion symmetry. For inversion symmetric materials, SFG is therefore interface specific.\cite{Shen2016} Several previous works have employed SFG to perform microscopy, but these approaches were limited to the mid-IR spectral region, where the small ratio of $\leq 4$ between IR and VIS wavelengths limits the achievable improvement of the spatial resolution.\cite{Raghunathan2011,Hanninen2017,Cimatu2006,Hoffmann2002} Our recent proof-of-concept experiment demonstrated SFG microscopy at IR and VIS wavelengths of ~12 $\mu$m and 0.5 $\mu$m, respectively, promising significantly improved spatial resolution. However, the spatial resolution in this approach was limited by the employed scanning focus method.\cite{Kiessling2019}

In this work, we achieve super-resolution in IR imaging using wide-field SFG microscopy. 
We employ a tunable IR free electron laser (IR-FEL) and a temporally synchronized VIS laser ($\lambda_{VIS} = 527$~nm) for SFG imaging in a home-built wide-field microscope. As a proof-of-concept, we image phonon polariton resonances in 4H-SiC sub-diffractional nanostructures.\cite{Caldwell2013,Razdolski2016,Kiessling2019,Lu2021} Using this approach, we demostrate a spatial resolution of $\sim\lambda_{IR}/9$. \\


IR and VIS laser beams illuminate the sample under an incidence angle of 50\,$^\circ$ and from the backside under normal incidence, respectively. The thus generated SFG light is collected by an objective and imaged on a camera, see Fig.\,\ref{fig1}c). As a mid-IR light source, we use the FEL installed at the Fritz Haber Institute, a high-power laser source offering a tunable wavelength in the mid- to far-IR range (3$\,\mu $m to 50$\,\mu $m) with a narrow bandwidth on the order of 0.3\,\%. 
We excite the sample with the IR light pulses at $\lambda_{IR}\approx11\,\mu$m with an energy of 5\,mJ in the macro pulse at 10\,Hz repetition rate and 9\,$\mu$J in the micro pulse.\cite{Schoellkopf2015} The visible light source is a frequency-doubled 1054\,nm table-top 55.5\,MHz femtosecond fiber oscillator (\textit{onefive Origami-HP}, pulse length 100\,fs) providing light at a wavelength of 527\,nm with $\sim$ 10\,nJ on the sample. Both beams are polarized parallel to the incidence plane and focused mildly to spotsizes of $\sim500\,\mu$m and $\sim200\,\mu$m, respectively, providing a high damage threshold. The temporal overlap of VIS and IR laser pulses is achieved by an RF-based synchronisation and verified by optical cross correlation using a phase-matched GaSe crystal. For more details on the timing and synchronisation system, see Ref. [\citenum{Kiessling2018}].

The SFG light generated in the sample is collected by the microscope setup consisting of a 50$\times$ long working distance objective (\textit{Mitutoyo M Plan Apo 50$\times$}, NA = 0.55, working distance wd = 13\,mm) and a 200\,mm tube lens. The axis of the objective is perpendicular to the sample surface. Behind the objective the VIS beam at $\lambda_{VIS}=527\,$nm is blocked by several filters (\textit{tunable bandpass 547/15, Semrock} and \textit{single-band bandpass 500/24, Semrock}). The SFG light at $\lambda_{SFG}\approx500\,$nm is detected with a high-sensitivity, electron multiplied CCD camera (\textit{PI-MAX 4, Teledyne Princeton Instruments}). For every image 100 FEL macro pulses are averaged corresponding to an aquisition time of 10\,s/image. The wide field approach provides a field of view of 275\,$\mu$m $\times$ 275\,$\mu$m with a 50x magnification.\\

In order to demonstrate the capabilities of the microscope, we image localized surface phonon polaritons (SPhP) in silicon carbide nanopillars (SiC NPs).\cite{Caldwell2013,Razdolski2016,Kiessling2019,Lu2021} SiC inhibits inherently a broken inversion symmetry. In the given geometry, only one unique  non-vanishing element $\chi^{(2)}_{xzx}=\chi^{(2)}_{yzy}$ of the second order nonlinear susceptibility $\chi^{(2)}$ tensor can contribute to the nonlinear signal,\cite{Paarmann2016} providing strong SFG intensity due to enhanced local optical fields at the polariton resonances.\cite{Razdolski2016} In particular, this geometry exclusively probes the normal-to-surface components of the local IR fields $E_{IR,z}$.\cite{Paarmann2016} In 4H-SiC nanostructures, localized SPhP\cite{Caldwell2015,Caldwell2013,Razdolski2016} are excited in the Reststrahlenband between the transversal optical (TO) and the longitudinal optical (LO) phonon frequencies 797\,cm$^{-1}$ and 964\,cm$^{-1}$, respectively, leading to a drastic enhancement of the local IR fields,\cite{Chen2014} and in consequence of the SFG emission.\cite{Kiessling2019,Razdolski2016} More specifically, SiC NPs arrays can support both monopolar and dipolar resonances, characterized by predominantly normal-to-surface and parallel-to-surface optical field enhancements, respectively.\cite{Caldwell2013,Chen2014} 

For this experiment, all the SiC nanopillar arrays are fabricated into semi-insulating 4H-SiC substrates using standard electron beam lithography (EBL) process. A 250\,nm PMMA 950 A4 layer was spin-coated at 3000\,rpm followed by deposition of 10\,nm-thick chromium as the conduction layer using thermal evaporation. The patterns were then defined using EBL followed by depositing 100\,nm-thick Nickel as a dry etch mask using electron beam evaporation. The SiC nanopillars were etched using reactive ion etching (RIE) using a mixture of O$_2$ and SF$_6$ at an etching rate of 375\,nm/min, and all the nanopillar structures are etched to 1\,$\mu$m.
Here, we show data based on NP square arrays from 1$\times$1 to 9$\times$9 pillars. Within each array, the NPs have a period of 5$\,\mu$m, pillar diameters of 1\,$\mu$m or 2$\,\mu$m, and a height of 1\,$\mu$m. An exemplary scanning electron microscopy (SEM) image of a 4$\times$4 NP array is shown in Fig.\,\ref{fig1}b).


\begin{figure*}[]
\centering
\includegraphics[scale=0.96]{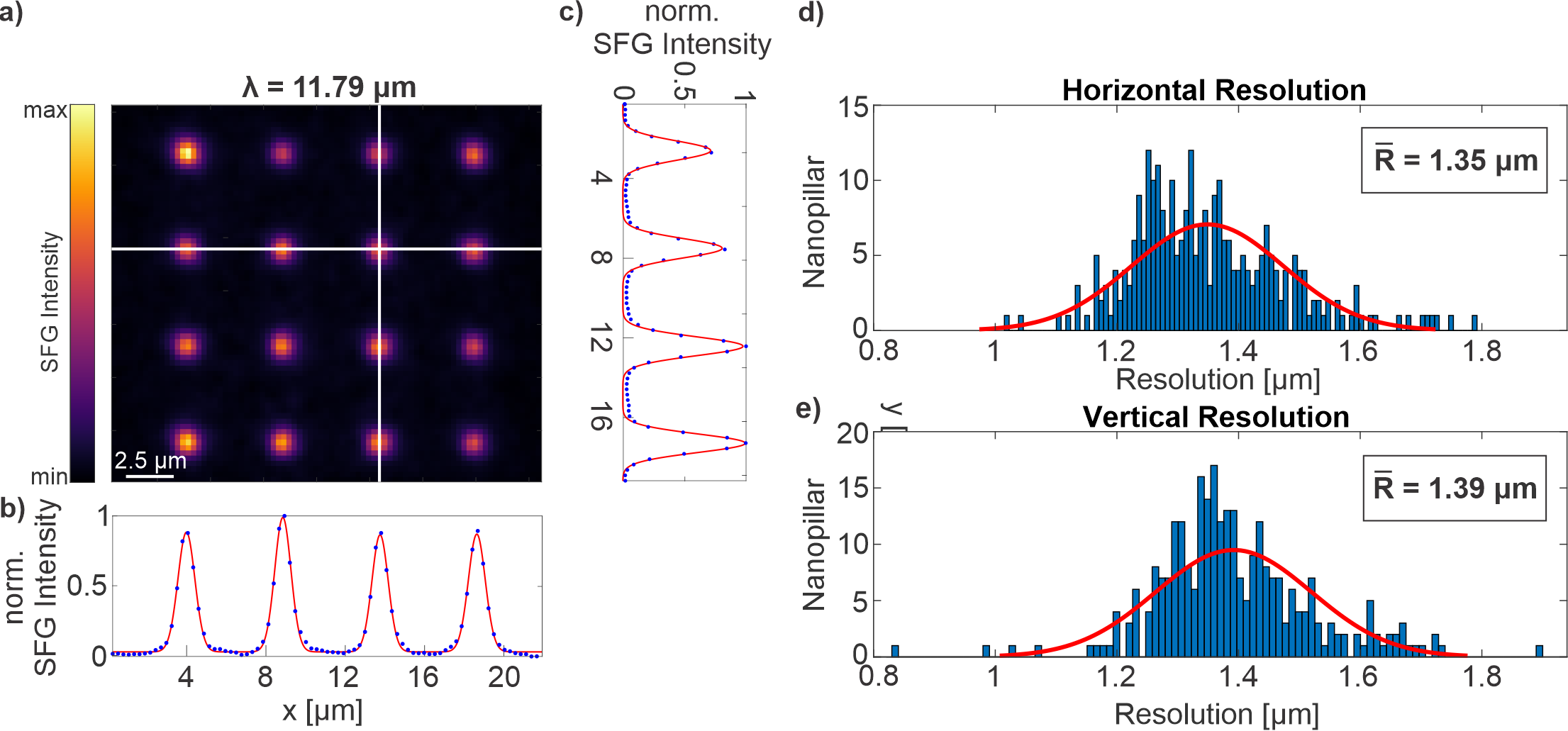} 
\caption{(a) SFG microscope image of the 4$\times$4 nanopillar array. The white lines indicate the position of the line profiles shown in (b) and (c). Blue dots show the measured light intensity and the solid red lines show fits by a series of double error functions, see Eq.\,\ref{eq:1}. (d) and (e) show histograms over all nanopillars (square arrays 1$\times$1 to 9$\times$9) for the vertical and horizontal resolution. A Gaussian fit gives a resolution of $1.35\,\mu $m and $1.39\,\mu $m for horizontal and vertical direction, respectively.}
\label{Fig:2}
\end{figure*}

Fig.\,\ref{Fig:2}a) shows an SFG image of a 4$\times$4 array of SiC NPs. We resolve the localized SPhP resonance in individual sub-diffractional NPs with sub-wavelength spacing. This provides the opportunity to estimate the spatial resolution of the microscope. For this purpose, we examine the images taken with the IR frequencies according to the monopole resonance of each nanopillar array, i.e. wavelengths between 11.6\,$\mu$m and 11.8\,$\mu$m.\cite{Caldwell2013} Specifically, we analyze horizontal and vertical line profiles, as exemplified by the white lines plotted in Fig.\,\ref{Fig:2}a). The corresponding line profiles are shown in Fig.\,\ref{Fig:2}b) \& c). Here, the blue dots show the measured SFG intensity, while the red solid lines show related double-error function fits according to 
\begin{multline}\label{eq:1}
I_{SFG}\left(x\right)=\Biggl[\sum_{i=1}^na_i*\biggl(erf\left(\frac{x+\mu_i+c_i}{\sigma_i*\sqrt{2}}\right)\\
*erf\left(\frac{x+\mu_i-c_i}{\sigma_i*\sqrt{2}}\right)+1\biggr)\Biggr]+b.
\end{multline}

 Here, $a$ is the intensity amplitude, $b$ is the total offset and $c$ the position of the error function. $\mu$ represents the center position of the pillar and $\sigma$ the standard deviation of the Gaussian broadening. The spatial resolution $R$ is calculated with $R=2.9*\sigma$.\cite{Engelbrecht2006} For statistical analysis, we imaged nanopillar square arrays of sizes n$\times$n, where n = 1...9. We therefore have fit parameters of 276 individual nanopillars with a diameter of $1\,\mu$m.  All fitted values for the horizontal and vertical resolution are plotted in histograms shown in Fig.\,\ref{Fig:2}d) \& e). The resulting normal distribution was fitted with a Gaussian (red line) leading to a horizontal resolution of $\left(1.35\pm0.13\right)\,\mu$m and a vertical resolution of $\left(1.39\pm0.13\right)\,\mu$m. 
   
With the microscope, we are able to resolve localized SPhP resonances in individual nanopillars with a size and spacing smaller than the resonant wavelength. These results show clearly that we achieve super resolution of the IR wavelength by a factor of $\sim\lambda_{IR}/9$. The spatial resolution for SFG microscopy is ultimately limited by the diffraction limit of the SFG light, $R_{Abbe}=\lambda_{SFG}/\left(2NA\right)=0.458\,\mu$m with $\lambda_{SFG}=504\,$nm and NA$=0.55$, which is about 1/3 of of the value achieved here. Since the mode volume of the localized SPhP is finite and on the order of the size of the NPs,\cite{Chen2014} the extracted value represents the convolution between the spatial extend of the mode and the resolution of the microscope. Therefore $R\approx1.4\,\mu$m is the upper limit of the resolution of the microscope. 

\begin{figure}[h!]
\centering
\includegraphics[scale=1]{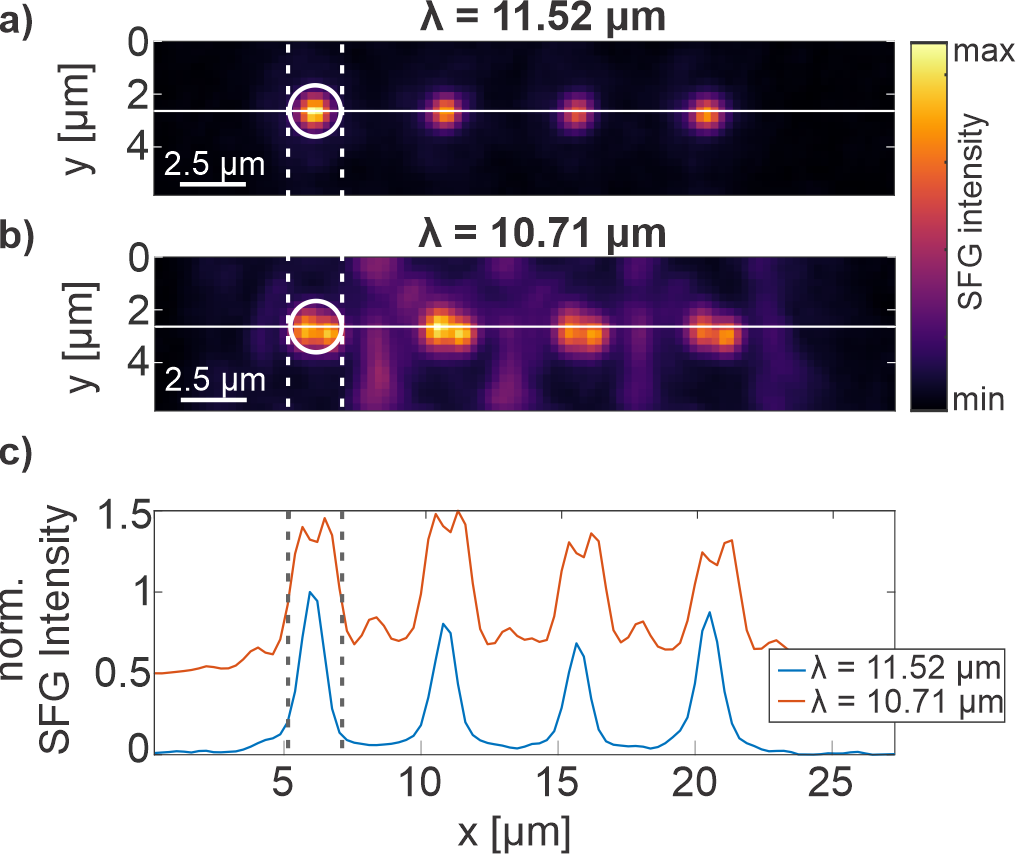} 
\caption{SFG images of the nanopillars resonant with the (a) monopole and (b) dipole mode.  The physical pillar size is indicated by the white circle. (c) Line profils along the white lines in a) and b) are plotted, showing the ability of resolving a two-peak-behavior on top of the nanopillar in the dipole mode in comparison to the monopole mode.  }
\label{fig3}
\end{figure}

A spatial resolution on the order of the size of the nanophotonic structures provides the opportunity to study the origin of the SPhP modes. In Fig.\,\ref{fig3}a) \& b), we present SFG images taken at two different IR wavelengths. The images were taken at the same sample position and show a line of four NPs with a diameter of 2$\,\mu$m. The white circles indicate the physical pillar size. In Fig.\,\ref{fig3}a) we observe a single peak of SFG intensity for each NP, while in Fig.\,\ref{fig3}b) we observe a double-peak. The different behavior is reflected in the line profiles, Fig.\,\ref{fig3}c), corresponding to the white lines in Fig.\,\ref{fig3}a) \& b), showing a single and a double-peak, respectively. This observation is consistent with monopole and dipole modes being excited at these two different wavelengths, shown in Fig.\,\ref{fig3}a) \& b), as expected from previous work.\cite{Chen2014,Gubbin2017} The images are in good agreement with simulations of the electric field distribution of SPhP in SiC NPs, showing that the monopole mode arises on the top of the nanopillars and the dipole mode arises from the edges of the nanopillars.\cite{Chen2014,Gubbin2017}


In summary, we demonstrate IR super resolution microscopy using sum frequency generation in a wide-field approach. We image individual sub-wavelength nanostructures with sub-diffractional spacing and achieve a resolution of a factor of $\sim9$ smaller than the wavelength of the probing IR light. With the super resolution SFG microscope we are able to identify monopole and dipole SPhP modes in individual SiC NPs. 
The spatial resolution of $R\approx 1.4\,\mu$m is three times larger than the theoretical Abbe limit, which is likely caused by the finite spatial extent of these modes.

In general, the wide-field microscope introduced here, opens up a wide range of applications. Beyond nanophotonic problems, our approach further allows to investigate spatially heterogeneous interfaces in electrochemistry\cite{Zwaschka2020} or domain structures in complex oxides\cite{Diebold2010}, by specifically using vibrational resonance imaging in these systems. In particular the large tunability of the IR-FEL further into the far-IR will allow to address many open questions in these areas, while preserving the high spatial resolution. \\

We acknowledge helpful discussion with Martin Thämer, Ben John, Tuhin Khan (all FHI Berlin). JDC acknowledges support from the U.S. Army Research Office (W911NF2110119), while GL was supported by a grant from NASA (80NSSC21K0766). Funding for GL was provided through an STTR program provided by the National Science Foundation, Division of Industrial Innovation and Partnerships (IIP) (Award No. 2014798). A portion of this research was conducted at the Vanderbilt Institute of Nanoscale Science and Engineering. 

\nocite{*}
\bibliography{SFG_mic}

\end{document}